\documentclass[epj]{svjour}
\usepackage{graphics}
\usepackage{mathtools}
\usepackage{comment}
\begin{document}
\title{Revisiting dynamics of interacting quintessence.}
\author{Patrocinio P\'erez \inst{1,\ast}, Ulises Nucamendi\inst{1,\dagger} \and Roberto De Arcia \inst{2,\ddagger}} 
%
%
\institute{Instituto de F\'isica y Matem\'aticas, Universidad Michoacana de San Nicol\'as de Hidalgo, Edificio C-3, Ciudad Universitaria, CP. 58040, Morelia, Michoac\'an, M\'exico. \and Departamento Ingenier\'ia Civil, Divisi\'on de Ingenier\'ia, Universidad de Guanajuato, C.P. 36000, Gto., M\'exico.}

\mail{\\ $\ast$ patrocinio.perez@umich.mx\\$\dagger$ unucamendi@gmail.com\\ $\ddagger$ robertodearcia@gmail.com}
\abstract{
We apply the tools of the dynamical system theory in order to revisit and uncover the structure of a nongravitational interaction between pressureless dark matter and dark energy described by a scalar field $\phi$. For a coupling function  $Q = -(\alpha d\rho_m/dt + \beta d\rho_\phi/dt )$, where t is the cosmic time, we have found that it can be  rewritten in the form $Q = 3H (\alpha \rho_m + \beta (d\phi/dt)^2 )/(1-\alpha +\beta)$, so that its dependence on the dark matter density  and on the kinetic term of the scalar field is linear and proportional to the Hubble parameter. We analyze the scenarios $\alpha=0$, $\alpha = \beta$ and $\alpha = -\beta$, separately and in order to describe the cosmological evolution we have calculated various observables. A notable result of this work is that, unlike for the noninteracting scalar field with exponential potential where five critical points appear, in the case studied here, with the exception of the matter dominated solution, the remaining singular points  are transformed into scaling solutions enriching the phase space. It is shown that for $\alpha \neq 0$, a separatrix arises modifying prominently the structure of the phase space.  This represents a novel feature no mentioned before in the literature.}
\PACS{
      {PACS-key}{Cosmology, dark matter, dark energy, quintessence, interacting quintessence model, dynamical system, phase space analysis, late-time scaling attractors.}
     } 
%
\maketitle
\section{Introduction}\label{sec-intro}
Recent cosmological observations indicate that our universe is currently undergoing an accelerated expansion phase. This has been confirmed by a wide variety of astronomical and cosmological data  which includes measurements of high red-shift supernovae Ia (SNIa) luminosity, temperature anisotropies of Cosmic Microwave Background (CMB), Baryon Acoustic Oscillations (BAO) and Large Scale Structure (LSS) among others \cite{Riess1998,Perlmutter1999,Hou2014,Anderson2014,Riess2019,Percival2004,Blake2011,Suzuki2012,An2018,Costa2019}.  To  explain  such  a  late  time acceleration in the context of general  relativity it is necessary to assume the existence of  a mysterious component  with  negative  pressure  broadly  known  as  dark  energy (DE). In the Lambda Cold Dark Matter ($\Lambda$CDM) model, the dark energy is described  by a cosmological constant with equation of state (EoS) parameter $\omega=-1$, which accounts approximately for the 70\% of the total energy content of the universe \cite{Weinberg1989,Carroll2001,Peebles2003}. It is still necessary to introduce an additional component dubbed Cold Dark Matter (CDM), which is postulated in order to increase the amount of structure formation needed to be in agreement with cosmological observations and represents around 25\% of the cosmic inventory . This component is typically associated to physics beyond the Standard Model of Particle physics (SM). Despite the standard cosmological model has successfully explained the observations it is not completely satisfactory from a theoretical point of view because it is plagued  by theoretical and philosophical problems at both the classical and quantum level such as the cosmic coincidence problem and  the vacuum energy problem  \cite{Zlatev1999,Avelino2016}. Moreover, as the accuracy of cosmological observations increases, tensions among different data sets have also emerged and this might be the first sign for physics beyond the $\Lambda$CDM model \cite{Verde2019,DiValentino2020,DiValentino2021}.\\

There are two main approaches one can follow in order to describe the observed universe acceleration: we either modify the gravity theory or we promote the cosmological constant to a dynamical dark energy. The interest in modified theories of gravity has significantly increased in the last years due to its ability to reproduce  a wide variety of astrophysical and cosmological observations.  According to the Lovelock theorem, GR represents  the  most  general  single metric theory that in four dimensions has field equations with at most second-order derivatives \cite{Lovelock1971}. Nonetheless, it may be extended in order to permit the field equations to be higher than second order, assuming the existence of dimensionality different from four or give up to locality \cite{Camanho2013,Crisostomi2018}. Among many alternatives, the \emph{scalar-tensor} theories of gravity represent  the prototypical way in which deviations from GR are modeled (see Refs. \cite{Clifton_review,Quiros_review,Bamba} for reviews). As an example, in \emph{Brans-Dicke} gravity one introduces an additional scalar mode besides the metric tensor replacing the gravitational coupling $G_N$ by a point-dependent scalar field \cite{Brans1961}. Alternatively, in the so-called $f(R,\mathcal{G})$ theories the Lagrangian is a general function of the Ricci scalar $R$ or the Gauss-Bonnet term $\mathcal{G}$ in the Jordan frame \cite{DeFelice2010}.  This gives rise to field equations with fourth-order derivatives and GR is recovered after the simplest choice of the function $f(R, \mathcal{G})\propto R$. As a consequence of introducing an arbitrary function there is a lot of freedom to explain the observed data.  Additionally, it is known that actions involving a finite number of power laws of curvature corrections and their derivatives can be considered as low-energy approximations to strings or supergravity theories giving rise to the so-called extended theories of gravity \cite{ext}. Finally, in the  Dvali-Gabadadze-Porrati (DGP) braneworld model one assumes the existence of a five-dimensional (5D) Minkowski spacetime of infinite volume within which ordinary four-dimensional (4D) Minkowski spacetime is embedded. It is precisely the presence of additional dimensions that realizes cosmic acceleration through the leakage of gravity into the extra-space at cosmological scales. This latter model, however, is plagued by ghost instabilities that cast doubts upon its validity \cite{DGP,Nicolis2004}.\\

Alternatively, the lack of knowledge on the nature of the dark sector has motivated several approaches to unveil their physical properties. One of the simplest scenarios is assuming the existence of a minimally coupled  scalar field $\phi$ with a self-interacting potential $V(\phi)$. This model arises from theories of gravity such as scalar-tensor theories and in the low-energy limit of string theories and has been the subject of interest due its ability to explaining various stages of the universe evolution \cite{a1,a2,a3,a4}. The canonical scalar field dubbed \emph{quintessence} resembles to the inflaton scalar field which was first proposed to explain the inflationary scenario which provides solutions to some issues of the big bang cosmology such as the initial singularity, flatness, horizon, homogeneity problems and the absence of magnetic monopoles \cite{Harko}. Compared to other scalar fields such as \emph{k-essence}, \emph{phantom} and \emph{quintom}, quintessence represents the simplest scenario without having theoretical problems such as the appearance of propagating ghost modes and Laplacian instabilities \cite{Tsujikawa}. Its dynamical behavior is characterized by the equation of state parameter $\omega_\phi = P_\phi /\rho_\phi$, where $P_\phi$ and $\rho_\phi$ denote its pressure and energy density respectively. For physically relevant cosmological scenarios the parameter is located into the interval $-1\leq \omega_\phi \leq -1/3$, where $\omega_\phi = -1$ corresponds to the cosmological constant model. Quintessence models can be classified in two classes, freezing and thawing, depending on whether the equation of state decreases towards $-1$ or departs from it \cite{Hara}.\\

There is also the possibility that dark energy might interact with dark matter through a nongravitational coupling $Q$ which is usually introduced at the level of the cosmological field equations. This represents an energy flow between the dark components and the sign of $Q$ determines the direction of the energy transfer: for
$Q > 0$ the matter fluid is giving energy to the dark energy fluid and vice versa for $Q < 0$. Notice that because of our current lack of knowledge about the nature of these two components, it would be imprudent to discard a nongravitational interaction between them.  Although this kind of models was first proposed in order to alleviate the cosmic coincidence problem, it was found that they also improve predictions on LSS, BAO, CMB anisotropies, galaxy clusters and $H(z)$ data among other cosmological and astrophysical experiments \cite{valid,Santos,18,Aljaf,Pan2020,Paliathanasis2018}. A wide variety of theoretical and phenomenological interacting scenarios have been proposed and investigated in the literature (see Ref. \cite{Wang_review,Bolotin_review} for reviews and references therein). To name a few, theoretical aspects such as the possibility to construct an interacting Lagrangian from which the interaction term can be derived is analyzed in \cite{bo}. In \cite{Banerjee} the authors study specific models of this class where they showed that cosmic chronometers and Type Ia supernovae data have a preference for interacting Quintessence models that lower $H_0$ relative to $ \Lambda$CDM. In \cite{gon} physical limits  on  the  equation of state parameter of the DE component non-minimally coupled  with  DM  are  examined  in  light  of the second law of thermodynamics and the positiveness of entropy. The study of the growth of cold dark matter density perturbations in the nonlinear regime is performed in \cite{Barros}, and in \cite{Linton} is shown that if the interaction between a quintessence field and cold dark matter is purely by momentum exchange, this generally leads to a dark energy sound speed that deviates from unity. Recently, assuming the dark  energy  component  as  a  quintessence  scalar  field  with Lagrangian function modified by the quadratic generalized uncertainty principle, in \cite{Paliathanasis2021} the authors investigate the behaviour of solutions of the field equations for some interacting models of special interests in the literature. Even though current cosmological data are compatible with such energy transfer models, the evidence so far is not completely conclusive \cite{costa,Yang2019}. \\

Since both the quintessence scalar field cosmology and the interacting dark energy models exhibit interesting phenomenological features, in the present work we perform a phase-space and stability analysis of the interacting scenario with exponential scalar potential and pressureless dark matter. Additionally, we compute some cosmological relevant quantities such as the dark energy density parameter, dark matter density parameter and the deceleration parameter. For the interacting term corresponding to a linear combination of the time derivatives of dark matter and scalar dark energy densities, $Q=-(\alpha \dot{\rho_m} + \beta \dot{\rho_{\phi}} )$, we analyzed the special cases depending on the value of the coupling parameters $\alpha=0$, $\alpha = \beta$ and $\alpha = -\beta$ separately. We will adopt dynamical system techniques which allow us to compute the equilibrium points  and we focus on the attractor solutions that can give rise to late time acceleration. If the attractor solution exists, the evolution of several models corresponding to a wide range of initial conditions converges towards an unique asymptotic behavior. \\

The structure of this work is as follows. In section \ref{sec:1} we review the basic equations governing the cosmology of the interacting dark sector scenario. In section \ref{sec:2} we introduce the master equations for the dynamical analysis and compute helpful cosmological parameters. We establish the interacting quintessence model and we discuss the adequate choice of the variables of the phase space. In section \ref{sec:3} the dynamical system is solved, the fixed points are determined and their respective stability is analyzed. We will draw conclusions and discuss future perspectives in section \ref{concl}. In appendix \ref{apen1} we report the regions of existence of the critical points for the different cases studied in this work and in appendix \ref{apen2} we will explicitly show the calculations around the conservation equations for the case of a nongravitational interacting scenario.

\section{Cosmological equations}

\label{sec:1}

In this section, we briefly introduce the dynamics of the cosmic components for a non gravitational interacting model. We consider only two components in the cosmic inventory: the quintessence scalar field representing the dark energy and the cold dark matter described by a pressureless barotropic perfect fluid. Let us assume a flat Friedmann-Lema\^{i}tre-Robertson-Walker(FLRW) metric with line element

\begin{equation}
ds^2 = -dt^2+ a(t)[dr^2+r^2 (d\theta^2 + \sin^2 \theta \, d\varphi^2)],
\end{equation}
where $a(t)$ is the scale factor and $t$ is the cosmic time. As a consequence of the interaction between the dark sector constituents the gravitational field equations become

\begin{eqnarray}\label{eq:friedmann1}
H^2 &\equiv & \left(\frac{\dot{a}}{a} \right) ^2 = \frac{\kappa^2}{3} \left( \rho_{m} + \rho_\phi \right), \\
\label{eq:friedmann2}
\dot{H} &=& -\frac{\kappa^2}{2} \left( \rho_{m} + \rho_\phi + P_\phi \right).
\end{eqnarray}
Here the dot represents the derivative with respect to the cosmic time, $\kappa^2 \equiv 8\pi G$, with $G$ the gravitational coupling constant,  $\phi$ is the scalar field, $\rho_{m}$ denotes the dark matter energy density and $\rho_\phi$ and $P_\phi$ represent the energy density and the pressure of the scalar field respectively
\begin{eqnarray}\label{eq:densidadphi}
\rho_\phi = \frac{1}{2} \dot{\phi}^2 + V(\phi), \\ \label{eq:presionphi}
P_\phi = \frac{1}{2}\dot{\phi}^2-V(\phi).
\end{eqnarray}
The function $V(\phi)$ is a self-interaction scalar potential and in this work we assume an exponential potential of the form $V(\phi) = V_0 \exp [-\kappa\lambda \phi]$, where $V_0>0$ is a constant with dimensions of mass and $\lambda$ is a dimensionless constant. This corresponds to the simplest example of quintessence scalar field and can be easily justified from high-energy phenomenology \cite{Ferreira1998}.
The cosmological dynamics of the exponential potential is captivating because of the appearance of accelerated solutions which can be employed to explain both the inflationary stage and the late time dynamics \cite{Lucchin1985,Geng2017}. Besides, the exponential potential has the interesting property of generating tracking solutions, i.e., for an appropriate choice of the parameter $\lambda$, the quintessence field evolves like radiation during the radiation-dominated era, and like matter during the matter-dominated era. This family of cosmological models have been extensively discussed in relation with early time inflation, high energy physics and late time accelerated scenarios \cite{Halliwell1987,Wands1993,Coley1997,Burd1988,Liddle1999,Wainwrightbook,Coleybook,Urena2012,Tamanini2014}.\\

Assuming the existence of an additional non gravitational interaction $Q$ which is introduced at the level of the cosmological field equations
\begin{eqnarray}\label{eq:conservationttm}
\dot{\rho}_\phi+3H(\rho_\phi+P_\phi) = Q, \\
\label{eq:conservationttphi}
\dot{\rho}_m+3H \rho_m = -Q,
\end{eqnarray}
where the sign of $Q$ determines the direction of the energy transfer: for $Q > 0$ the matter fluid is giving energy to the dark energy fluid and vice versa for $Q<0$. It is important to mention that, in order to satisfy the local energy conservation, equations \eqref{eq:conservationttm} and  \eqref{eq:conservationttphi} are not independent due to the Bianchi identities as we show in appendix \ref{apen2}. In this work we will restrict the discussion to quintessence models where dark energy is assumed to be a scalar field with self-interacting exponential potential, however, we can assume the existence of more complicated potential terms \cite{Gonzalez2006,Leon2009,Morris2013,Hossain2014,Tzanni2014,Bahamonde2018}. For dynamical systems studies of interacting dark energy as a perfect fluid we can find in the literature \cite{Olivares2008,Quartin2008,Caldera2009,Quercellini2010,Li2010}.

Finally, the evolution for the scalar field is given by
\begin{eqnarray}
\ddot{\phi} + 3H\dot{\phi} + \frac{d V(\phi)}{d \phi} = \frac{Q}{\dot{\phi}}. \label{kg}
\end{eqnarray}

\section{Phase space variables} \label{sec:2}
The application of the dynamical system theory is specially useful when one deals with scalar-field cosmological models \cite{Faraoni2013,Fadragas2014,Garcia2015,Cid2016}. It should be mentioned that, as far as we know, reference \cite{Belinskii1985} is a pioneering study on the application of dynamics systems to cosmology. From the dynamical systems tools one may obtain very useful information on the asymptotic dynamics of the system which is characterized by: i) source critical points which may be pictured as past attractors, ii) saddle equilibrium configurations that attract the phase space orbits in one direction but repel them in another
direction, iii) attractor solutions to which the system evolves for a wide range of initial conditions, or iv) limit cycles, among others. \\

In order to trade the system of second order equations \eqref{eq:conservationttphi} and \eqref{kg} by a system of autonomous ordinary differential equations one has to choose a suitable set of variables. In general, there are many possible ways to achieve this task. The most common one is to consider the normalized variables introduced in \cite{wands}:
\begin{equation}\label{eq:definiciones}
X^2 \equiv \frac{\kappa^2 \dot{\phi}^2}{6H^2}\mbox{,}\phantom{0.0} Y^2 \equiv \frac{\kappa^2 V(\phi)}{3H^2},\phantom{0.0} \lambda \equiv -\frac{1}{\kappa V}\left(\frac{\partial V}{\partial \phi}\right).
\end{equation}
Here we are assuming that only expanding cosmologies arise: $H\geq 0$ (with $Y\geq 0$). The constraint (\ref{eq:friedmann1}) written in terms of the set of normalized variables takes the form
\begin{equation}\label{eq:FriedmannConstraint}
1=\Omega_m+X^2+Y^2= \Omega_m+\Omega_\phi,
\end{equation}
where $\Omega_m$ and $\Omega_\phi$ are the density parameters usually defined for the dark matter and scalar field respectively. Hence, the physically meaningful phase space corresponds to the  region
\begin{equation}\label{eq:restricciones}
\Psi = \{(X,Y): 0 \leq Y \leq 1,\,\,  0\leq X^2+Y^2\leq 1\}.
\end{equation}
Using this set of variables \eqref{eq:definiciones} and the Friedmann constraint \eqref{eq:FriedmannConstraint} the system of equations which governs the dynamics reduces to the following set of autonomous equations
\begin{equation}\label{eq:Xequation}
X'= -3X+\sqrt{\frac{3}{2}}\lambda Y^2 -X\left(\frac{\dot{H}}{H^2} \right) +\frac{\kappa^2 }{6H^3}\left(\frac{Q}{X} \right),
\end{equation}
\begin{equation}\label{eq:Yequation}
Y' = -Y\left[\sqrt{\frac{3}{2}}\lambda X + \frac{\dot{H}}{H^2} \right],
\end{equation}
where the prime denotes the derivative with respect to the logarithm of the scale factor. It is important to mention that, in general, the system of equations \eqref{eq:Xequation} - \eqref{eq:Yequation} is not closed unless the coupling function $Q$ can be expressed in terms of the variables \eqref{eq:definiciones}.
In this work, we consider the  coupling function
\begin{equation}\label{eq:interaction}
Q =-( \alpha \dot{\rho_m}+\beta \dot{\rho_\phi}),
\end{equation}
which was incorrectly studied in \cite{Shahalam2015} and \cite{Binayak} as we show at the end of this section. Replacing (\ref{eq:interaction}) in  (\ref{eq:conservationttm}) - (\ref{eq:conservationttphi}) and after some algebra, the system of equations takes the form
\begin{eqnarray}\label{eq:system1}
& (1-\alpha) \dot{\rho}_m - \beta \dot{\rho}_\phi = -3H {\rho}_m, \\
\label{eq:system2}
& \alpha \dot{\rho}_m + (1+\beta)\dot{\rho}_\phi = -3H \dot{\phi}^2.
\end{eqnarray}
We can algebraically solve for $\dot{\rho}_\phi$ and $\dot{\rho}_m$
\begin{eqnarray}\label{eq:derscadensity}
\dot{\rho}_\phi = 3H \left[\frac{\alpha \rho_m - (1-\alpha) \dot{\phi}^2}{1+\beta-\alpha}\right], \\
\label{eq:dermatdensity}
\dot{\rho}_m = -3H \left[\frac{(1+\beta) \rho_m + \beta \dot{\phi}^2}{1+\beta-\alpha}\right] ,
\end{eqnarray}
and  we can write the interaction term (\ref{eq:interaction}) into the form
\begin{equation}\label{eq:finalinteraction}
Q = 3H \left[\frac{\alpha \rho_m + \beta \dot{\phi}^2}{1+\beta-\alpha}\right].
\end{equation}
Then we have shown that the interaction function (\ref{eq:interaction}) is equivalent to an interaction term $Q$ lineally proportional to the Hubble parameter and a lineal combination of the dark matter density and the kinetic term of the scalar field.

In order to proceed to the phase-space analysis it is necessary to compute the functions $\dot{H}/H^2$ and $Q$ in terms of variables $X$ and $Y$. From equation (\ref{eq:friedmann2}) in addition  with (\ref{eq:densidadphi}) and (\ref{eq:presionphi}), we obtain
\begin{equation}\label{eq:friedmann2b}
\frac{\dot{H}}{H^2}= -\frac{\kappa^2}{2H^2}\left(\rho_m+\dot{\phi}^2 \right)=-\left( \frac{3}{2}\right)\left(1+X^2-Y^2 \right).
\end{equation}
Finally, with the help of \eqref{eq:definiciones}, (\ref{eq:finalinteraction}), (\ref{eq:friedmann2b}), we derive the first order dynamical system from equations (\ref{eq:Xequation}) and (\ref{eq:Yequation}) as
\begin{eqnarray}\nonumber
X' &=& -\frac{3(1-\alpha)}{(1-\alpha+\beta)}X+\sqrt{\frac{3}{2}}\lambda Y^2\\ \nonumber
   & &+\left(\frac{3}{2} \right) X \left[ 1+X^2-Y^2 \right]\\  \label{eq:xprim}
   & & +\left( \frac{3}{2}\right) \frac{\alpha}{(1-\alpha+\beta)}\frac{(1-X^2-Y^2)}{X},\\ \label{eq:yprim}
Y' &=& -Y\left[\sqrt{\frac{3}{2}}\lambda X -\left(\frac{3}{2}\right)(1+X^2-Y^2) \right].
\end{eqnarray}
The choice $\alpha = \beta = 0$ represents the non-interaction scenario studied in  \cite{wands}. For the case $\beta=0$ it can be shown that making the transformation $\frac{3\alpha}{1-\alpha}\rightarrow \alpha $ we can reproduce the model II analyzed in \cite{Bohmer2008}.

The deceleration parameter which is defined as one of the geometrical parameters through which the dynamics of the universe can be quantified is depicted by
\begin{equation}
q \equiv -\frac{a\ddot{a}}{\dot{a}^2} = -\frac{\dot{H}}{H^2}-1=\frac{1}{2}+\left( \frac{3}{2}\right)\left( X^2-Y^2\right),
\end{equation}
while the effective equation of state parameter  $\omega_{eff}\equiv\frac{P_{tot}}{\rho_{tot}}$ can be written as
\begin{equation}
\omega_{eff} = X^2-Y^2.
\end{equation}

Finally, we notice that, in the case $\alpha \neq 0$ and the limit $X \rightarrow 0$, $Y \rightarrow Y_{0}$ with $Y_{0}\neq \pm 1$, the last term of the right hand side of \eqref{eq:xprim} diverges and therefore the system of equations \eqref{eq:xprim}-\eqref{eq:yprim} does not satisfy the fundamental existence and uniqueness theorem for nonlinear Ordinary Differential Equations Systems because it is not continuously differentiable (See page 74 of the reference  \cite{Perko}). Actually, for $Y \rightarrow Y_{0}$ with $Y_{0}\neq \pm 1$, we have the limits

\begin{equation}\label{eq:lim+}
\lim_{X \rightarrow 0^{+}} \frac{dY}{dX} = \left \{ \begin{matrix} 0^{+} & \mbox{if } \frac{\alpha}{(1-\alpha+\beta)}>0 \mbox{,}
\\
\\
0^{-} & \mbox{if }\frac{\alpha}{(1-\alpha+\beta)}<0\mbox{,}\end{matrix}\right.
\end{equation}

\begin{equation}\label{eq:lim-}
\lim_{X \rightarrow 0^{-}} \frac{dY}{dX} = \left \{ \begin{matrix} 0^{-} & \mbox{if } \frac{\alpha}{(1-\alpha+\beta)}>0 \mbox{,}
\\
\\
0^{+} & \mbox{if }\frac{\alpha}{(1-\alpha+\beta)}<0\mbox{,}\end{matrix}\right.
\end{equation}
this shows that the vertical line $X=0$ is a separatrix in the compact phase space \eqref{eq:restricciones}, namely, the dynamics of the region $X>0$ is completely disconnected causally from the region $X<0$. Actually, in a neighbourhood of $X=0$ there exist two different trajectories with the same initial or end condition:
in the case $\alpha/(1-\alpha+\beta)>0$, the point $X=0, Y=Y_{0}$ (with $Y_{0}\neq \pm 1$) is the initial condition for two different trajectories which depart from it (for example, see bottom panels of Figs. \ref{fig:phase-space-alpha=beta} and \ref{fig:phase-space-alpha=-beta}; by the contrary, in the case $\alpha/(1-\alpha+\beta)<0$, the point $X=0, Y=Y_{0}$ (with $Y_{0}\neq \pm 1$) is the end point for two different trajectories (for example, see top panels of Figure \ref{fig:phase-space-alpha=beta} and Figure \ref{fig:phase-space-alpha=-beta}.

As it was previously mentioned, the interaction coupling (\ref{eq:interaction}) was studied in references \cite{Shahalam2015}, \cite{Binayak} where a corresponding mistaken dynamical system was analyzed providing wrong results in the critical points found and the subsequent stability analysis; specifically, the first of equations (9) of reference \cite{Shahalam2015} and equation (18) of reference \cite{Binayak} are wrong because, in both, the interaction term is missing (which provides precisely the term that is proportional to the inverse of the variable X in the right hand side of equation \eqref{eq:xprim} in this work).

\section{Dynamical analysis, critical points, and stability}
\label{sec:3}
This section is devoted to analyze the cosmological dynamics of the system of cosmological equations  \eqref{eq:conservationttphi} and \eqref{kg} by the system of autonomous ordinary differential equations (ODE-s) \eqref{eq:xprim} and \eqref{eq:yprim} in the form $\bf{\dot{x}} = \bf{f}(\bf{x})$. Here $\bf{x}$  is called a point in the phase space  and $\bf{f}$ corresponds to the column vector of the autonomous equations. A critical (or equilibrium) point  $\bf{x}_c$, is a point in the phase space that satisfies the condition $\bf{f}(\bf{x}_c) = \bf{0}$.  In order to determine the stability  properties of the system we expand around $\bf{x}_c$ as $\bf{x} = \bf{x_c}+\bf{u}$, with $\bf{u}$ the column vector of the perturbations. Therefore, for each critical point we expand the perturbation equations up to first order as $ \dot{\bf{u}} = I\!\!M \bf{u}$, where the matrix $I\!\!M$ contains the coefficients of the perturbation equations. Finally, the eigenvalues of $I\!\!M$ are evaluated for all critical points in order to determine its type and stability.\\

For the dynamical system defined by the equations (\ref{eq:xprim}) and (\ref{eq:yprim}), there are six critical points, these are reported in Table \ref{tab:1}.
\begin{table*}
\caption{Critical points of the dynamical system.}
\label{tab:1}       
\begin{center}
\begin{tabular}{lll}
\hline\noalign{\smallskip}
Point &$X$ & $Y$ \\
\noalign{\smallskip}\hline\noalign{\smallskip}
$A_+$ &$\sqrt{\left[ \frac{(1-\beta)+ \sqrt{\Delta(\alpha,\beta)}}{2(1+\beta-\alpha)} \right]}$ & $0$ \\
$A_-$ &$-\sqrt{\left[ \frac{(1-\beta)+ \sqrt{\Delta(\alpha,\beta)}}{2(1+\beta-\alpha)} \right]}$ & 0 \\
$B_+$ &$\sqrt{\left[ \frac{(1-\beta)- \sqrt{\Delta(\alpha,\beta)}}{2(1+\beta-\alpha)} \right]}$ & 0 \\
$B_-$ &$-\sqrt{\left[ \frac{(1-\beta)- \sqrt{\Delta(\alpha,\beta)}}{2(1+\beta-\alpha)} \right]}$ & 0 \\
$C_+$ &$\frac{(3+(1+\beta-\alpha)\lambda^2) + \sqrt{\Gamma(\alpha,\beta, \lambda) } }{2\sqrt{6} \lambda (1+\beta-\alpha)}$ &$ \frac{1}{6}\sqrt{ -3\lambda^2+\frac{9(3+\sqrt{\Gamma(\alpha, \beta, \lambda)})}{(1-\alpha+\beta)^2\lambda^2}+\frac{3(6-12\alpha+6\beta+\sqrt{\Gamma(\alpha, \beta, \lambda)})}{(1-\alpha+\beta)}}$ \\
$C_-$ &$\frac{(3+(1+\beta-\alpha)\lambda^2) - \sqrt{\Gamma(\alpha,\beta, \lambda) } }{2\sqrt{6} \lambda (1+\beta-\alpha)}$ &$ \frac{1}{2\sqrt{3}}\sqrt{\frac{6-12\alpha+6\beta+\sqrt{\Gamma(\alpha, \beta, \lambda)}}{(1-\alpha+\beta)}-\frac{3(-3+\sqrt{\Gamma(\alpha,\beta, \lambda)})}{(1-\alpha+\beta)^2\lambda^2} -\lambda^2}$ \\
\noalign{\smallskip}\hline
\end{tabular}
\end{center}
\end{table*}
We have defined the following equations:
\begin{equation}
\Delta(\alpha, \beta)\equiv (1-\beta)^2 -4\alpha(1-\alpha+\beta),
\end{equation}
\begin{equation}
\Gamma(\alpha, \beta, \lambda) \equiv (1-\alpha+\beta)\lambda^2 [\lambda^2 -6 (1+2\beta)]+9.
\end{equation}

To determine the existence of critical points $A_+, A_-, B_+$ and $B_-$ we use the constraint given by equation (\ref{eq:restricciones}). Note that the $X$-component for the critical points $C_+$ and $C_-$ is antisymmetric under $\lambda \rightarrow -\lambda$, this is $X(\alpha, \beta, -\lambda) = - X(\alpha, \beta, \lambda)$ while the second restriction $0\leq X^2+Y^2\leq 1$ holds. For the critical points $A_+, A_-, B_+$ and $B_-$ it is easy to find their stability since there is no dependence on the parameter $\lambda$. On the other hand, for the critical points $C_+$ and $C_-$, this represent a more complicated task and for this reason we analyze only some special cases in order to simplify the analysis.

\subsection{Scenario $\alpha = 0$}

\begin{figure}
\resizebox{0.45\textwidth}{!}{%
  \includegraphics{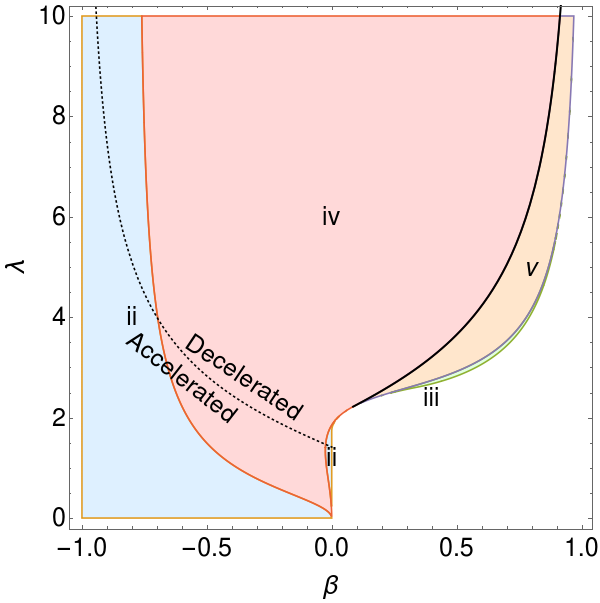} }
\caption{Part of the region of existence of solution of the critical point $E_-$. The region ii contains stable points, iii unstable points, iv spiral stable points, v spiral unstable points and center points on black line. The black dotted line divides accelerated and decelerated critical points.}
\label{fig:2}
\end{figure}

\begin{table*}
\caption{Critical points of the dynamical system for $\alpha =0 $.}
\label{tab:2}
\begin{center}
\begin{tabular}{llllll}
\hline\noalign{\smallskip}
Point &$X$ & $Y$ & $\Omega_m$&$\Omega_\phi$&$\frac{\Omega_m}{\Omega_\phi}$  \\
\noalign{\smallskip}\hline\noalign{\smallskip}
$O$ & $0$ & $0$& 1& 0& undefined\\
$D_+$ &$\sqrt{ \frac{1-\beta}{1+\beta} }$ & 0 & $\frac{2\beta}{1+\beta}$& $\frac{1-\beta}{1+\beta}$&$\frac{2\beta}{1-\beta}$\\
$D_-$ &$-\sqrt{ \frac{1-\beta}{1+\beta} }$ & 0 & $\frac{2\beta}{1+\beta}$& $\frac{1-\beta}{1+\beta}$&$\frac{2\beta}{1-\beta}$\\
$E_+$ &$\frac{3+(1+\beta)\lambda^2 + \sqrt{\Gamma_1} }{2\sqrt{6} \lambda (1+\beta)}$ &$ \frac{1}{2\sqrt{3}}\sqrt{\frac{3(3+\sqrt{\Gamma_1})}{(1+\beta)^2\lambda^2}+\frac{(6+6\beta-\sqrt{\Gamma_1})}{(1+\beta)}-\lambda^2}$ &$\frac{\lambda^2(1+\beta)(1+2\beta)-\sqrt{\Gamma_1}-3}{2\lambda^2(1+\beta)^2}$&$\frac{3+(1+\beta)\lambda^2+\sqrt{\Gamma_1}}{2\lambda^2(1+\beta)^2}$&$\frac{(1+\beta)\lambda^2-\sqrt{\Gamma_1}-3}{6}$\\
$E_-$ &$\frac{3+(1+\beta)\lambda^2 - \sqrt{\Gamma_1 } }{2\sqrt{6} \lambda (1+\beta)}$ &$ \frac{1}{2\sqrt{3}}\sqrt{\frac{3(3-\sqrt{\Gamma_1})}{(1+\beta)^2\lambda^2}+\frac{(6+6\beta+\sqrt{\Gamma_1})}{(1+\beta)}-\lambda^2}$&$\frac{\lambda^2(1+\beta)(1+2\beta)+\sqrt{\Gamma_1}-3}{2\lambda^2(1+\beta)^2}$&$\frac{3+(1+\beta)\lambda^2-\sqrt{\Gamma_1}}{2\lambda^2(1+\beta)^2}$&$\frac{(1+\beta)\lambda^2+\sqrt{\Gamma_1}-3}{6}$ \\
\noalign{\smallskip}\hline
\end{tabular}
\end{center}
\end{table*}
For a vanishing coupling constant $\alpha$, the interacting kernel reduces to
\begin{equation}
Q = -\beta\dot{\rho}_\phi = \frac{3H \beta}{1+\beta}\dot{\phi}^2,
\end{equation}
while the functions $\Delta$ and $\Gamma$ take the form
\begin{equation}
\Delta(0, \beta)\equiv (1-\beta)^2,
\end{equation}
\begin{equation}
\Gamma_1 \equiv \Gamma(0, \beta, \lambda) \equiv (1+\beta)^2 \lambda^4-6\lambda^2 (1 +\beta)(1+2\beta)+9.
\end{equation}
In this scenario we have found five critical points (the critical points $A_+, A_-, B_+$ and $B_-$ reported in Table \ref{tab:1}, reduce to $O, D_+$ and $D_-$ ) reported in Table \ref{tab:2}. The existence conditions, stability, acceleration, and $\omega_{eff}$ are reported in Table \ref{tab:3}. The region of existence of the critical point $E_+$ is the region 1 reported in appendix A and we show this in Figure  \ref{fig:1}. The region of existence of the critical point $E_-$ is region 2 reported in appendix A and this region is not bounded in the parameter $\lambda$, for this reason, we only report part of this and we show in Figure \ref{fig:2} the region considered.\\

Here, $q<0$ means that there is acceleration and we can see in Table \ref{tab:3} that only for the critical point $E_-$ we have acceleration and this is shown in Figure \ref{fig:2}, where below the black dotted curve we have acceleration and in the other case we have deceleration.
\begin{figure}
\resizebox{0.45\textwidth}{!}{%
  \includegraphics{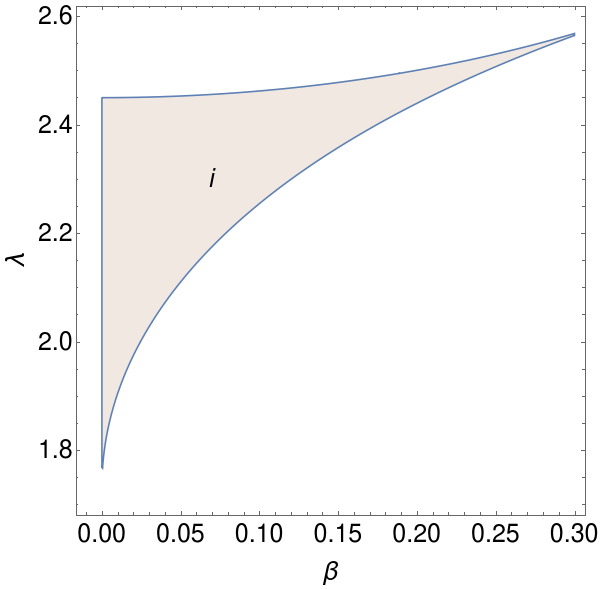} }
\caption{Region of existence for the critical point $E_+$. They are decelerated saddle points in all region i.}
\label{fig:1}
\end{figure}

\begin{table*}
\caption{The physically meaningful critical points of the autonomous system for the case $\alpha = 0$.}
\label{tab:3}       
\begin{center}
\begin{tabular}{lllll}
\hline\noalign{\smallskip}
Point & Existence &  Stability  & $q<0$ & $\omega_{eff}$  \\
\noalign{\smallskip}\hline\noalign{\smallskip}
O &All $\lambda$ and & Saddle for all $\lambda$ and $-1<\beta<1$ &No & 0\\
  & all $\beta$ with $\beta \neq -1$  & Unstable for all $\lambda$ and ($\beta>1$ or $\beta <-1$) & ($q=\frac{1}{2}$) & \\
$D_+$ & All $\lambda$ and & Saddle for $\lambda >  \sqrt{\frac{6}{1-\beta^2}}$ and $0\leq \beta<1$ & No & $\frac{1-\beta}{1+\beta} $ \\
 & $0\leq \beta<1$  & Unstable for $\lambda < \sqrt{\frac{6}{1-\beta^2}}$  and $0\leq \beta<1$&  & \\
 $D_-$ & All $\lambda$ and  & Saddle for $\lambda <-\sqrt{\frac{6}{1-\beta^2}}$ and $0\leq \beta <1 $ & No & $\frac{1-\beta}{1+\beta} $\\
 &  $0\leq \beta<1$ &  Unstable for $\lambda >-\sqrt{\frac{6}{1-\beta^2}}$ and $0\leq \beta <1$ &  & \\
$E_+$ &  Region 1 & Saddle in all region i in Figure \ref{fig:1} & No& $\frac{\lambda^2-3+(\lambda^2-6)\beta+\sqrt{\Gamma_1}}{6(1+\beta)}$  \\
 & (in the appendix)  & & & \\
$E_-$  & Region 2 & Stable: area ii in Figure \ref{fig:2} &$-1<\beta \leq 0$ & $\frac{\lambda^2-3+\beta(\lambda^2-6)-\sqrt{\Gamma_1}}{6(1+\beta)}$\\
 & (in the appendix)& Unstable: area iii in Figure \ref{fig:2} & and & \\
  &  & Stable spiral: area iv in Figure \ref{fig:2} & $0<\lambda<\sqrt{\frac{2-4\beta}{1+\beta}}$ & \\
 & & Unstable spiral: area v in Figure \ref{fig:2} & &\\
 & & Centre: black curve in Figure \ref{fig:2} & &\\
\noalign{\smallskip}\hline
\end{tabular}
\end{center}
\end{table*}

\begin{figure}[h]
\resizebox{.5\textwidth}{!}{%
  \includegraphics{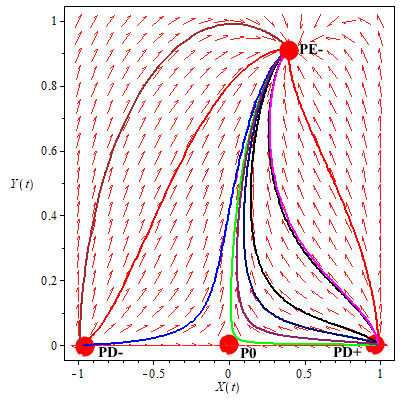}
  \includegraphics{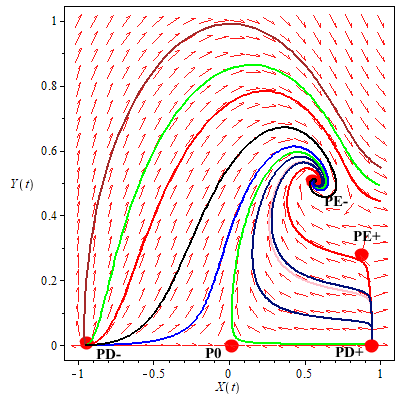} }
\caption{Phase portrait of the dynamical system \eqref{eq:xprim} and \eqref{eq:yprim} for the case $\alpha=0$ and the specific choice $\beta = 0.01,\lambda=1$ (left panel) and $\beta =0.05,\lambda=2.3$ (right panel). In the first scenario, Point $D_+$ and Point $D_-$ correspond to unstable nodes, Point $O$ is saddle and Point $E_-$ describes an accelerated stable solution which can be of cosmological interest. For the latter scenario, Point $D_+$ and Point $D_-$ are related again with unstable nodes, Point $O$ is saddle, Point $E_+$ (absent in the previous case) is saddle and Point $E_-$ is a stable solution. This scenario is decelerated and it is not favored by observations. It is important to mention that the different colors of the trajectories are only for illustrative purposes.}
\label{fig:phase-space-alpha=0}
\end{figure}

The critical points of the dynamical system for the choice $\alpha= 0$, as well as their stability properties, are listed and briefly discussed below. For a couple of illustrative scenarios see Figure \ref{fig:phase-space-alpha=0}.\\

(i) Point $O$: The matter dominated solution exists for a coupling constant $\beta \neq -1$ and it is independent of the specific form of the self-interacting potential. Here the effective equation of state parameter vanished $(\omega_{eff}=0)$ and therefore there is no acceleration ($q=1/2$). For $-1<\beta <1$ this point behaves as saddle, otherwise it is unstable. In the non-interacting scenario this solution behaves always as saddle, therefore the chance that this point can be related to an origin of some trajectories in the phase space is due to the  presence of a nongravitational interaction.

(ii) Point $D_+$: The dark energy scaling solution exists for all values of the parameter $\lambda$ and for $0\leq \beta<1$. The special case $\beta= 0$ denotes an universe dominated by the scalar field kinetic energy ($\Omega_m= 0$, $X= 1$ and $Y=0$) and the limit $\beta\rightarrow 1$ corresponds to a matter dominated universe ($\Omega_m\rightarrow 1$, $X\rightarrow 0$ and $Y=0$). The effective EoS parameter is depicted by $\omega_{eff}=\frac{1-\beta}{1+\beta}\in (0, 1]$, and then the decelerated parameter is non-negative corresponding to a decelerated solution. For $\lambda<\sqrt{\frac{6}{1-\beta^2}}$ the solution is a past attractor and for  $\lambda>\sqrt{\frac{6}{1-\beta^2}}$ it behaves as saddle and therefore it cannot be a late-time state of the universe. \\

(iii) Point $D_-$: The scaling solution does not depend on the parameter $\lambda$, but it is still required that $0\leq \beta<1$. The limit $\beta = 0$ corresponds to a stiff matter universe ($\Omega_m= 0$, $X= -1$ and $Y=0$), and $\beta\rightarrow 1$ denotes a matter dominated universe ($\Omega_m\rightarrow 1$, $X\rightarrow 0$ and $Y=0$). The equation of state parameter is always positive $\omega_{eff}=\frac{1-\beta}{1+\beta}\in (0, 1]$, therefore the solution is decelerated. This point behaves as saddle for $\lambda<-\sqrt{\frac{6}{1-\beta^2}}$ and it is an unstable node if $\lambda>-\sqrt{\frac{6}{1-\beta^2}}$.  \\

(iv) Point $E_+$: Exists for the region 1 reported in the appendix. It behaves always as saddle and thus it cannot attract the universe at late times. For $\beta = 0$ the matter energy density vanished $(\Omega_m=0)$ while $X= \frac{\lambda}{\sqrt{6}}$ and $Y= \sqrt{1-\frac{\lambda^2}{6}}$, which corresponds to the scalar field dominated universe. For $\beta = \frac{1}{3}$ and $\lambda = \sqrt{\frac{27}{4}}$ the universe has the components $\Omega_m= \frac{1}{2}$, $X=\frac{1}{2}$ and $Y=0$. In general, $\Omega_m$, $X$, and $Y$ never vanish simultaneously, despite this, the scalar field kinetic energy never dominates. \\

(v) Point $E_-$: The scaling solution exists for the region 2 reported in the appendix. This point can be unstable (node and spiral), a centre or either stable (node and spiral). This solution is accelerated if the parameters lie in the region $\lambda < \sqrt{\frac{2-4\beta}{(1+\beta)}}$ and $-1<\beta\leq 0$ and then it can be the late-time state of the universe. As an example, for $\beta = -\frac{1}{2}$ and $\lambda = 1$ we have the following quantities $X\approx 0.187$, $Y\approx 0.939$, $\Omega_m\approx 0.082$ and $q= -0.270$. In the non-interacting scenario this point is either a stable node or a stable spiral.

\subsection{Scenario $\alpha = \beta $}
\begin{figure}
\resizebox{0.45\textwidth}{!}{%
  \includegraphics{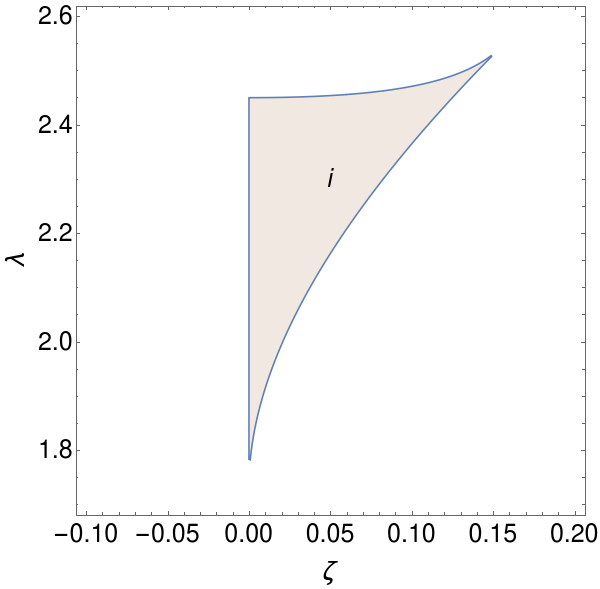} }
\caption{Region of existence of solution $H_+$ when $\alpha = \beta \equiv \zeta$. It contains saddle points.}
\label{fig:3}
\end{figure}

\begin{figure}
\resizebox{0.45\textwidth}{!}{%
  \includegraphics{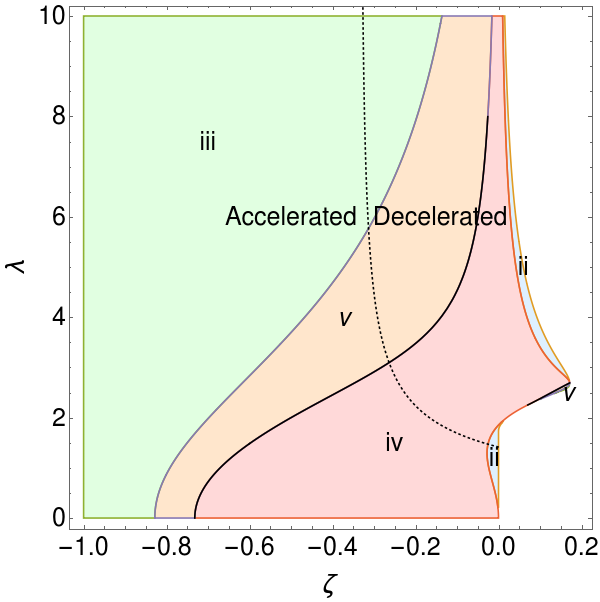} }
\caption{Part of the region of existence of the critical point $H_-$ when $\alpha = \beta \equiv \zeta$. Region ii contains stable points, iii unstable points, iv spiral stable points, v spiral unstable points and center points are located on the black lines. The black dotted curve divides the accelerated and decelerated universe.}
\label{fig:4}
\end{figure}

\begin{table*}
\caption{Critical points of the dynamical system for $\alpha = \beta\equiv \zeta $.}
\label{tab:4}
\begin{center}
\begin{tabular}{llllll}
\hline\noalign{\smallskip}
Point &$X$ & $Y$ & $\Omega_m$&$\Omega_\phi$&$\frac{\Omega_m}{\Omega_\phi}$  \\
\noalign{\smallskip}\hline\noalign{\smallskip}
$F_+$ &$\sqrt{\left[ \frac{1-\zeta+ \sqrt{\zeta^2-6\zeta+1}}{2} \right]}$ & $0$&$\frac{1+\zeta-\sqrt{\zeta^2-6\zeta+1}}{2}$ &$ \frac{1-\zeta+\sqrt{\zeta^2-6\zeta+1}}{2}$ &$\frac{1+\zeta-\sqrt{\zeta^2-6\zeta+1}}{1-\zeta+\sqrt{\zeta^2-6\zeta+1}}$ \\
$F_-$ &$-\sqrt{\left[ \frac{1-\zeta+ \sqrt{\zeta^2-6\zeta+1}}{2} \right]}$ & $0$&$\frac{1+\zeta-\sqrt{\zeta^2-6\zeta+1}}{2}$ &$ \frac{1-\zeta+\sqrt{\zeta^2-6\zeta+1}}{2}$ &$\frac{1+\zeta-\sqrt{\zeta^2-6\zeta+1}}{1-\zeta+\sqrt{\zeta^2-6\zeta+1}}$ \\
$G_+$ &$\sqrt{\left[ \frac{1-\zeta- \sqrt{\zeta^2-6\zeta+1}}{2} \right]}$ & 0 & $ \frac{1+\zeta+\sqrt{\zeta^2-6\zeta+1}}{2}$& $\frac{1-\zeta-\sqrt{\zeta^2-6\zeta+1}}{2}$ & $\frac{1+\zeta+\sqrt{\zeta^2-6\zeta+1}}{1-\zeta-\sqrt{\zeta^2-6\zeta+1}}$\\
$G_-$ &$-\sqrt{\left[ \frac{1-\zeta- \sqrt{\zeta^2-6\zeta+1}}{2} \right]}$ & 0 & $ \frac{1+\zeta+\sqrt{\zeta^2-6\zeta+1}}{2}$& $\frac{1-\zeta-\sqrt{\zeta^2-6\zeta+1}}{2}$ & $\frac{1+\zeta+\sqrt{\zeta^2-6\zeta+1}}{1-\zeta-\sqrt{\zeta^2-6\zeta+1}}$\\
$H_+$ &$\frac{(3+\lambda^2) + \sqrt{\Gamma_2 }}{2\sqrt{6} \lambda }$ &$ \sqrt{\frac{6(1-\zeta)-\lambda^2-\sqrt{\Gamma_2}}{12}+\frac{3+\sqrt{\Gamma_2}}{4\lambda^2}}$ & $\frac{(1+2\zeta)\lambda^2-3-\sqrt{\Gamma_2}}{2\lambda^2}$ & $\frac{3+(1-2\zeta)\lambda^2+\sqrt{\Gamma_2}}{2\lambda^2}$ & $ \frac{(1-2\zeta^2)\lambda^2-3-\sqrt{\Gamma_2}}{6+2(\zeta-1)\zeta \lambda^2}$\\
$H_-$ &$\frac{(3+\lambda^2) - \sqrt{\Gamma_2 } }{2\sqrt{6} \lambda }$  &$ \sqrt{\frac{6(1-\zeta)-\lambda^2+\sqrt{\Gamma_2}}{12}+\frac{3-\sqrt{\Gamma_2}}{4\lambda^2}}$ & $ \frac{(1+2\zeta)\lambda^2-3+\sqrt{\Gamma_2}}{2\lambda^2}$ & $\frac{3+(1-2\zeta)\lambda^2-\sqrt{\Gamma_2}}{2\lambda^2}$ & $ \frac{(1-2\zeta^2)\lambda^2-3+\sqrt{\Gamma_2}}{6+2(\zeta-1)\zeta \lambda^2 }$\\
\noalign{\smallskip}\hline
\end{tabular}
\end{center}
\end{table*}
In this case, the interaction is
\begin{equation}
Q =-\zeta(\dot{\rho}_m +\dot{\rho}_\phi)= 3H\zeta(\rho_m+\dot{\phi}^2),
\end{equation}
where $\alpha=\beta\equiv \zeta $. The critical points are reported in Table \ref{tab:4}, and the $\Delta$ and $\Gamma$ functions are:
\begin{eqnarray}\nonumber
&& \Delta(\zeta, \zeta) = \zeta^2-6\zeta+1 \\ \nonumber
&& \Gamma_2 \equiv \Gamma(\zeta, \zeta, \lambda) = \lambda^4-6\lambda^2(1+2\zeta)+9
\end{eqnarray}
The existence, stability, acceleration $(q<0)$ and $\omega_{eff}$ are reported in Table \ref{tab:5}. The region of existence of the critical points $H_+$ and $H_-$ are called region 3 and region 4 respectively, this is reported in the appendix A. The region of existence of the critical point $H_+$ is shown in Figure \ref{fig:3}. For the critical point $H_-$, the region of existence is not bounded in $\lambda$, for this reason, the stability of the region is shown in Figure \ref{fig:4}.

\begin{table*}
\caption{The physically meaningful critical points of the autonomous system for the case $\alpha = \beta \equiv \zeta$.}
\label{tab:5}       
\begin{center}
\begin{tabular}{lllll}
\hline\noalign{\smallskip}
Point & Existence &  Stability  & $q<0$ & $\omega_{eff}$  \\
\noalign{\smallskip}\hline\noalign{\smallskip}
$F_+$  & $ 0\leq \zeta < 3-2\sqrt{2}$ & Saddle for $\lambda > \sqrt{6}$ and $\zeta=0$ & No & $\frac{(1-\zeta)+ \sqrt{\zeta^2-6\zeta+1}}{2} $ \\	
  & & Saddle for $\lambda >w_1(\zeta)$ and $ 0<\zeta < 3-2\sqrt{2}$ &  &  \\
 &  & Unstable for $\lambda < \sqrt{6}$ and $\zeta=0$ & & \\	
 &  & Unstable for $\lambda <w_1(\zeta)$ and $ 0<\zeta < 3-2\sqrt{2}$ & & \\
$F_-$  & $0\leq \zeta < 3-2\sqrt{2}$ & Saddle for $\lambda <- \sqrt{6}$ and $\zeta = 0$ & No &  $\frac{(1-\zeta)+ \sqrt{(1-\zeta)^2-4\zeta}}{2}$ \\	
  &  & Saddle for $\lambda < -w_1(\zeta)$ and $ 0<\zeta < 3-2\sqrt{2}$ &  &  \\
  &  & Unstable for $\lambda > -\sqrt{6}$ and $ \zeta=0$ & & \\	
  &  & Unstable for $\lambda > -w_1(\zeta)$ and $ 0<\zeta < 3-2\sqrt{2}$ & & \\
$G_+$ & $ 0\leq \zeta < 3-2\sqrt{2}$ & Saddle for all $\lambda$ and $ \zeta=0$ &  No & $\frac{(1-\zeta)- \sqrt{(1-\zeta)^2-4\zeta}}{2} $\\	
 &  & Saddle for $\lambda < w_2(\zeta)$ and $ 0<\zeta < 3-2\sqrt{2}$ &  & \\
  &  & Stable for $\lambda > w_2(\zeta)$ and $ 0<\zeta < 3-2\sqrt{2}$ & & \\
$G_-$ & $ 0\leq \zeta < 3-2\sqrt{2}$ & Saddle for all $\lambda$ and $ \zeta=0$ & No & $\frac{(1-\zeta)- \sqrt{(1-\zeta)^2-4\zeta}}{2} $ \\	
 &  & Saddle for $\zeta > -w_2(\zeta)$ and $ 0<\zeta < 3-2\sqrt{2}$ &  &  \\
  &  & Stable for $\lambda < -w_2(\zeta)$ and $ 0<\zeta < 3-2\sqrt{2}$ & &\\
 $H_+$ & Region 3  & Saddle $\sqrt{3}<\lambda<\sqrt{6}$ and $\zeta = 0$ & No & $\frac{\lambda^2-3+\sqrt{\Gamma_2}}{6}$ \\
  & (in the appendix) & Saddle point in region i in Figure \ref{fig:3} with $\zeta\neq 0$ & & \\
$H_-$ & Region 4 & Stable: part ii in Figure \ref{fig:4} & $(\lambda<\sqrt{\frac{2}{1+3\zeta}}$ and $-\frac{1}{3}<\zeta\leq 0)$ & $\frac{\lambda^2-3-\sqrt{\Gamma_2}}{6}$ \\
 &  (in the appendix)& Unstable: part iii in Figure \ref{fig:4} & or & \\
 &  & Stable spiral: part iv in \ref{fig:4} & ($\lambda> 0$ and $-1<\zeta\leq-\frac{1}{3}$)           &  \\
 &  & Unstable spiral: part v in \ref{fig:4} & &\\
 &  & Centre: Black curve in Figure \ref{fig:4} & & \\
\noalign{\smallskip}\hline
\end{tabular}
\end{center}
\end{table*}
For Table \ref{tab:7} we define $w_1(\zeta)$ and $w_2(\zeta)$:
\begin{equation} \nonumber
w_1(\zeta)=\sqrt{\frac{12\zeta+3-3\zeta^2-3\sqrt{(-1+\zeta)^2(1-6\zeta+\zeta^2)}}{4\zeta}},
\end{equation}
\begin{equation}\nonumber
w_2(\zeta) = \sqrt{\frac{12\zeta+3-3\zeta^2+3\sqrt{(-1+\zeta)^2(1-6\zeta+\zeta^2)}}{4\zeta}}.
\end{equation}

\begin{figure}[h]
\resizebox{.5\textwidth}{!}{%
  \includegraphics{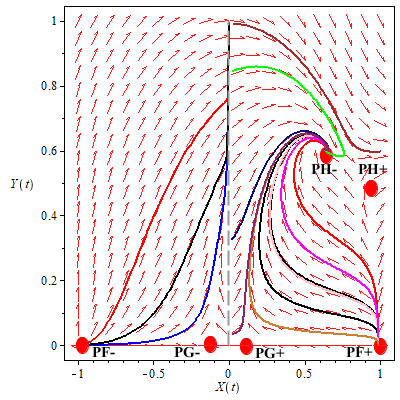}
  \includegraphics{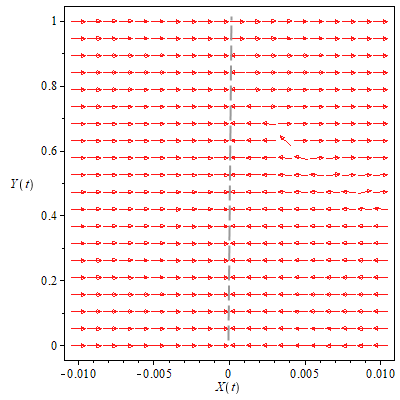} }
\resizebox{.5\textwidth}{!}{%
    \includegraphics{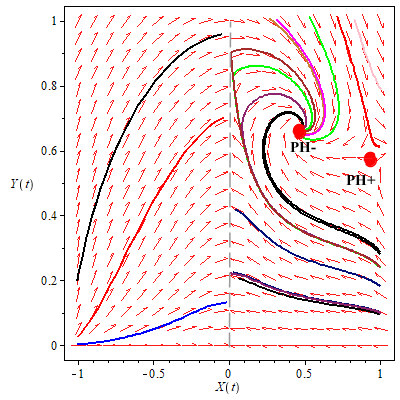}
      \includegraphics{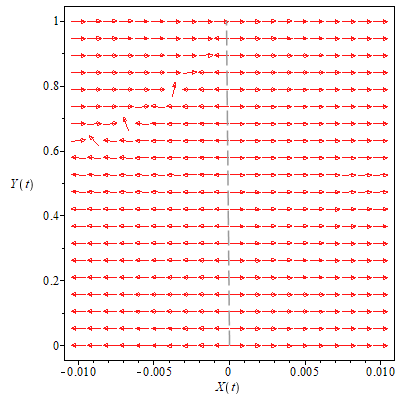} }
\caption{Phase portrait of the autonomous system of ODE \eqref{eq:xprim} and \eqref{eq:yprim} for the case $\alpha=\beta$ and the specific choice $(\beta = -0.01,\lambda=2)$ -top panels- and $(\beta =0.-5,\lambda=2.5)$ - bottom panels-. Here the gray line denotes the separatrix $X=0$ which exists whenever $Y \neq 1$ and divides the regions where the kinetic term of the scalar field is positive or negative. It can be seen that the trajectories end at it for $\alpha/(1-\alpha+\beta)<0$ or depart from it if $\alpha/(1-\alpha+\beta)>0$. Due to an optical illusion one could assume that some trajectories can cross the separatrix, and then the behavior of the trajectories around the separatrix is shown in the right-hand side. In the first scenario, Point $F_+$ and Point $F_-$ denote unstable nodes, Point $G_+$ and Point $G_-$ correspond to saddle nodes, Point $H_+$ describes an accelerated stable solution, which can be of cosmological interest. For the latter scenario, Point $F_+$, Point $F_-$, Point $G_+$ and Point $G_-$ vanish, leaving Point $H_+$ describing a saddle node and Point $H_-$ which is associated to a stable spiral. This scenario is decelerated.}
\label{fig:phase-space-alpha=beta}
\end{figure}

The critical points of the dynamical system for the choice $\alpha= \beta$, as well as their stability properties, are listed and briefly discussed below. For a couple of illustrative scenarios see Figure \ref{fig:phase-space-alpha=beta}.\\

(i) Point $F_+$: Exists for all $\lambda$ and $0\leq \zeta < 3-2\sqrt{2}$, this is a scaling solution for $\zeta\neq 0$. When $\zeta=0$, it is dominated by the scalar field kinetic energy $\Omega_m =0$, $X=1$ and $Y=0$, on the other hand when $\zeta\rightarrow 3-2\sqrt{2}$ we have the following components $\Omega_m\rightarrow 0.58$, $X\rightarrow 0.64$ and $Y=0$. For all $\lambda$ and $\zeta =0$ or for $\lambda>w_1(\zeta)$ and $0<\zeta<3-2\sqrt{2}$ this is a saddle point. For $\lambda<\sqrt{6}$ and $\zeta=0$ or for $\lambda < w_1(\zeta)$ and $0<\zeta<3-3\sqrt{2}$ this is an unstable point. On the other hand $\omega_{eff}\in (0.41, 1]$, hence, there is no acceleration.\\

(ii) Point $F_-$: This solution corresponds to a decelerated universe and exists only if $0\leq \zeta < 3-2\sqrt{2}$. For $\zeta\neq 0$, we retrieve the scaling solution and thus it can alleviate the cosmic coincidence problem. For the uncoupled scenario $\zeta=0$, the universe is dominated by the scalar field kinetic energy $\Omega_m =0$, $X=-1$ and $Y=0$. In the limit case $\zeta\rightarrow 3-2\sqrt{2}$ the cosmological parameters are depicted by $\Omega_m\rightarrow 0.58$, $X\rightarrow -0.64$ and $Y=0$. The stability properties are listed in Table \eqref{tab:1}.\\

(iii) Point $G_+$: Exists for all $\lambda$ and $0\leq \zeta < 3-2\sqrt{2}$, this is scaling solution for $\zeta\neq 0$. When $\zeta =0$ we have $\Omega_m=1$, $X=0$, $Y=0$, this corresponds to a matter dominated universe. For $\zeta \rightarrow 3-2\sqrt{2} $ the components are $\Omega\rightarrow 0.58$, $X\rightarrow 0.64$ and $Y=0$, we see that for this critical point the scalar field never dominates. $\omega_{eff}\in [0, 0.41)$, hence, there is no acceleration. It is a saddle point for all $\lambda$ and $\zeta =0$ or for $\lambda<w_2(\zeta)$ and $0< \zeta < 3-2\sqrt{2}$. The critical point is stable for $\lambda>w_2(\zeta)$ and $0< \zeta < 3-2\sqrt{2}$.\\

(iv) Point $G_-$: This solution exists for all values of the parameter $\lambda$, and  $0\leq \zeta < 3-2\sqrt{2}$ is required. The case $\zeta\neq 0$ corresponds to  a scaling solution, while $\zeta =0$ is related to a matter dominated solution $\Omega_m=1$, where the dynamical variables $X$ and $Y$ identically vanished. The cosmological parameters read  $\Omega_m\rightarrow 0.58$, $X\rightarrow -0.64$ and $Y=0$ in the limit $\zeta \rightarrow 3-2\sqrt{2}$, therefore the scalar field never dominates. The equation of state parameter is located into the interval $\omega_{eff}\in [0, 0.41)$, and hence the solution is always decelerated. \\

(v) Point $H_+$: Exists for the region 3 reported in the appendix A and shown in Figure \ref{fig:3}. It is a saddle point on all region. The particular case for $\lambda=\sqrt{6}$ and $\zeta=0$ has the components $\Omega_m$, $X=1$, and $Y=0$, this one is dominated by the scalar field kinetic energy; for $\lambda=\sqrt{3}$ and $\zeta=0$ the values of the components are $\Omega_m=0$, $X=\frac{1}{2}$ and $Y=\frac{1}{2}$, this is dominated by the scalar field; for $\lambda= \sqrt{6}$ and $\zeta= 0.1$ the values of the components are $\Omega_m\approx 0.23$, $X\approx 0.68$ and $Y\approx 0.01$. For all this region does not have accelerated universes.  \\

(vi) Point $H_-$: Exists for the region 4 reported in the appendix, this is not bounded on the parameter $\lambda$, this region is shown in Figure \ref{fig:4}. We see from Figure \ref{fig:4} that exist saddle point, stable point, unstable point, spiral stable point, spiral unstable point and centre point for this critical point depending on the values of $\lambda$ and $\zeta$. The black dashed lines divide the accelerated and decelerated regions. For $\lambda = 4$ and $\zeta= -0.4$ we have $\Omega_m\approx 0.49$, $X\approx 0.16$, $Y\approx 0.68$, $\omega_{eff}\approx -0.44$ and $q\approx -0.16$, this point is an accelerated unstable spiral. For $\lambda= 4$, $\zeta=-0.2$ we have $\Omega_m\approx 0.88$, $X=Y\approx 0.23$, $\Omega_{eff}=0$ and $q=0.5$, this is a decelerated spiral point.\\

\subsection{Scenario $\alpha = -\beta$}
\begin{table*}
\caption{Critical points of the dynamical system for $\alpha = -\beta \equiv \eta$.}
\label{tab:6}
\begin{center}
\begin{tabular}{llllll}
\hline\noalign{\smallskip}
Point &$X$ & $Y$ & $\Omega_m$&$\Omega_\phi$&$\frac{\Omega_m}{\Omega_\phi}$  \\
\noalign{\smallskip}\hline\noalign{\smallskip}
$I_+$ &$\sqrt{\left[ \frac{1+\eta+ \sqrt{9\eta^2-2\eta+1}}{2(1-2\eta)} \right]}$ & $0$ & $\frac{1-5\eta-\sqrt{9\eta^2-2\eta+1}}{2(1-2\eta)} $ & $\frac{1+\eta+\sqrt{9\eta^2-2\eta+1}}{2(1-2\eta)} $ & $\frac{1-\eta-\sqrt{9\eta^2-2\eta+1}}{2\eta} $ \\
$I_-$ & $-\sqrt{\left[ \frac{1+\eta+ \sqrt{9\eta^2-2\eta+1}}{2(1-2\eta)} \right]}$ & 0 & $\frac{1-5\eta -\sqrt{9\eta^2-2\eta+1}}{2(1-2\eta)} $ & $\frac{1+\eta+\sqrt{9\eta^2-2\eta+1}}{ 2(1-2\eta)} $ & $\frac{1-\eta-\sqrt{9\eta^2-2\eta+1}}{2\eta}$\\
$J_+$ & $\sqrt{\left[ \frac{1+\eta- \sqrt{9\eta^2-2\eta+1}}{2(1-2\eta)} \right]}$ & 0 &$ \frac{1-5\eta+\sqrt{9\eta^2-2\eta+1}}{2(1-2\eta)}$ & $\frac{1+\eta-\sqrt{9\eta^2-2\eta+1}}{2(1-2\eta)}$ & $\frac{1-\eta+\sqrt{9\eta^2-2\eta+1}}{2\eta}$\\
$J_-$ &$-\sqrt{\left[ \frac{1+\eta- \sqrt{9\eta^2-2\eta+1}}{2(1-2\eta)} \right]}$ & 0 & $\frac{1-5\eta+\sqrt{9\eta^2-2\eta+1}}{2(1-2\eta)}$ & $ \frac{1+\eta-\sqrt{9\eta^2-2\eta+1}}{2(1-2\eta)}$ & $ \frac{1-\eta+\sqrt{9\eta^2-2\eta+1}}{2\eta}$\\
$K_+$ &$\frac{3+(1-2\eta)\lambda^2 + \sqrt{\Gamma_3 } }{2\sqrt{6} \lambda (1-2\eta)}$ &$\sqrt{\frac{3(3+\sqrt{\Gamma_3})}{12\lambda^2(1-2\eta)^2}+\frac{6-18\eta-\sqrt{\Gamma_3}}{12(1-2\eta)}-\frac{\lambda^2}{12}} $ &$\frac{(1-2\eta)^2\lambda^2-3-\sqrt{\Gamma_3}}{2(1-2\eta)^2\lambda^2} $ & $ \frac{3+(1-2\eta)^2\lambda^2+\sqrt{\Gamma_3}}{ 2(1-2\eta)^2\lambda^2}$ & $\frac{(1+2(\eta-1)\eta)\lambda^2-3-\sqrt{\Gamma_3}}{6+2(\eta-1)\eta \lambda^2} $\\
$K_-$ &$\frac{3+(1-2\eta)\lambda^2 - \sqrt{\Gamma_3 } }{2\sqrt{6} \lambda (1-2\eta)}$ &$\sqrt{\frac{3(3-\sqrt{\Gamma_3})}{12\lambda^2(1-2\eta)^2}+\frac{6-18\eta+\sqrt{\Gamma_3}}{12(1-2\Gamma)}-\frac{\lambda^2}{12}} $ & $\frac{(1-2\eta)^2\lambda^2-3+\sqrt{\Gamma}}{2(1-2\eta)^2\lambda^2} $ & $\frac{(1-2\eta)^2\lambda^2+3-\sqrt{\Gamma_3}}{2(1-2\eta)^2\lambda^2} $ & $\frac{(1+2\eta(\eta-1))\lambda^2-3+\sqrt{\Gamma_3}}{6+2(\eta-1)\eta \lambda^2} $\\
\noalign{\smallskip}\hline
\end{tabular}
\end{center}
\end{table*}
For this case the interaction is
\begin{equation}
Q= \eta(-\dot{\rho}_m+\dot{\rho}_\phi)= \frac{3H\eta }{1-2\eta}\left( \rho_m-\dot{\phi}^2\right),
\end{equation}
where $ \alpha=-\beta \equiv \eta$, same interaction was studied in \cite{Binayak} but again with wrong equations. The critical points are reported in Table \ref{tab:6}. The $\Delta$ and $\Gamma$ functions are:
\begin{eqnarray}
&&\Delta(\eta, -\eta)= 9\eta^2-2\eta +1, \\
&&\Gamma_2\equiv \Gamma(\eta, -\eta, \lambda)=(1-2\eta)^2\lambda^2(\lambda^2-6)+9.
\end{eqnarray}
The existence, stability, acceleration $(q<0)$ and $\omega_{eff}$ are reported in Table \ref{tab:7}. The region of existence of the critical points $K_+$ and $K_-$ are reported in the appendix A label by region 5 and region 6 respectively.

\begin{figure}
\resizebox{0.45\textwidth}{!}{%
  \includegraphics{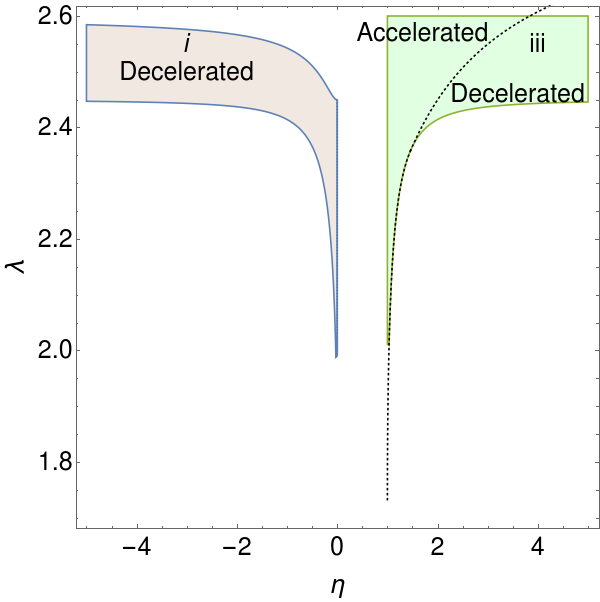} }
\caption{Part of the region of existence of solution for the critical point $K_+$ when $\alpha = -\beta \equiv \eta $. The region labeled by i contains saddle points and iii unstable points.}
\label{fig:5}
\end{figure}
\begin{figure}
\resizebox{0.45\textwidth}{!}{%
  \includegraphics{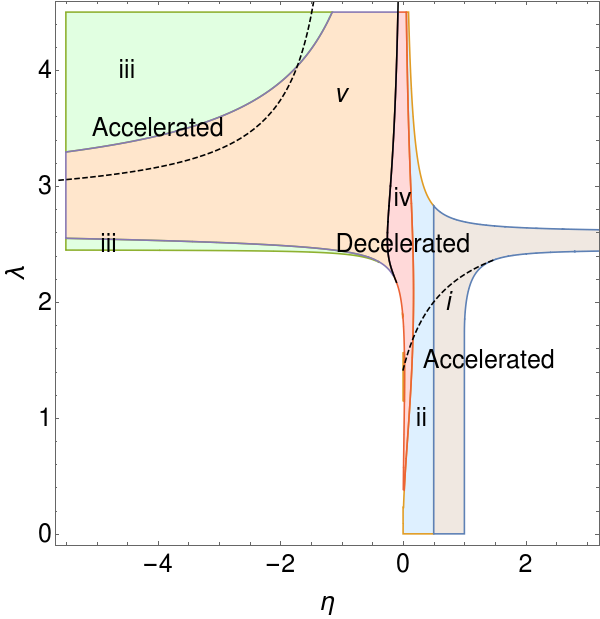} }
\caption{Part of the region of existence of solution for the critical point $K_-$ when $\alpha = -\beta\equiv \eta$. The region labeled by i contains saddle points, ii stable points, iii unstable points, iv spiral stable points, v spiral unstable points and center points on black line. The black dotted curves divide the accelerated and decelerated universe.}
\label{fig:6}
\end{figure}
\begin{table*}
\caption{The physically meaningful critical points of the autonomous system for the case $\alpha = -\beta \equiv \eta$.}
\label{tab:7}
\begin{center}
\begin{tabular}{lllll}
\hline\noalign{\smallskip}
Point & Existence &  Stability  & $q<0$ & $\omega_{eff}$  \\
\noalign{\smallskip}\hline\noalign{\smallskip}
$I_+$  & $\eta \leq 0$ & Saddle for $\lambda >\sqrt{6}$ and $ \eta = 0$ & No & $\frac{1+\eta+\sqrt{9\eta^2-2\eta+1}}{2(1-2\eta)} $ \\
  &  & Saddle for $\lambda > f_2(\eta)$ and $ \eta < 0$ & &  \\
   &  & Unstable for $\lambda < \sqrt{6}$ and $\eta=0$ &  & \\	
		&  & Unstable for $\lambda < f_2(\eta)$ and $\eta<0$ & & \\	
$I_-$  & $\eta \leq 0$ & Saddle for $\lambda < -\sqrt{6}$ and $\eta = 0$ & No & $\frac{1+\eta+\sqrt{9\eta^2-2\eta+1}}{2(1-2\eta)}$\\
  &  & Saddle for $\lambda < -f_2(\eta)$ and $\eta < 0$ &  &  \\
  &  & Unstable for $\lambda > -\sqrt{6}$ and $\eta = 0$ & & \\
  &  &  Unstable for $\lambda > -f_2(\eta)$ and  $\eta < 0$ & & \\
$J_+$  & $\eta \geq 0$ & Saddle for all $\lambda$ and $\eta=0$ & No & $\frac{1+\eta-\sqrt{9\eta^2-2\eta+1}}{2(1-2\eta)}$\\
  & with $\eta \neq \frac{1}{2}$ & Saddle for $\lambda < f_1(\eta)$ and  $0<\eta\leq \frac{1}{3}$ &  & \\
  &  & Saddle for $\lambda < f_2(\eta)$ and $\frac{1}{3}<\eta<\frac{1}{2}$  & & \\
  &  & Saddle for $\lambda > f_2(\eta)$ and $\eta>\frac{1}{2}$ & & \\
  &  & Stable for $\lambda > f_1(\eta)$ and  $0<\eta\leq \frac{1}{3}$ & & \\
  &  & Stable for $\lambda > f_2(\eta$ and $\frac{1}{3}<\eta<\frac{1}{2}$ & &\\
  &  & Unstable for $\lambda < f_2(\eta)$ and  $\eta>\frac{1}{2}$ & & \\
$J_-$ &$\eta \geq 0$  & Saddle for all $\lambda$ and $\eta=0$ & No & $ \frac{1+\eta-\sqrt{9\eta^2-2\eta+1}}{2(1-2\eta)}$\\
  & with $\eta \neq \frac{1}{2}$ & Saddle for $\lambda > -f_1(\eta)$ and $0<\eta\leq \frac{1}{3}$ & & \\
  &  & Saddle for $\lambda > -f_2(\eta)$ and $\frac{1}{3}<\eta<\frac{1}{2}$ & & \\
  &  & Saddle for $\lambda < -f_2(\eta)$ and  $\eta>\frac{1}{2}$ &  &  \\
  &  & Stable for $\lambda < -f_1(\eta)$ and $0<\eta\leq \frac{1}{3}$ & & \\
  &  & Stable for $\lambda < -f_2(\eta)$ and $\frac{1}{3}<\eta<\frac{1}{2}$ & & \\
  &  & Unstable for $\lambda > -f_2(\eta)$ and  $\eta>\frac{1}{2}$ & & \\
 $K_+$ & Region 5  & Saddle for $f_3(\eta)<\lambda<f_2(\eta)$ and $\eta \leq 0$,  region i in Figure \ref{fig:5} & See Figure \ref{fig:5}. & $\frac{\lambda^2-3-2\eta(\lambda^2-6)+\sqrt{\Gamma_3}}{6(1-2\eta)}$ \\
 & (in the appendix)  &  Unstable for $f_3(\eta)<\lambda$ and $\eta > 1$, region iii in Figure \ref{fig:5} &  & \\
 $K_-$ & Region 6  & Saddle: part i in Figure \ref{fig:6} & See Figure \ref{fig:6} & $\frac{\lambda^2-3-2\eta(\lambda^2-6)-\sqrt{\Gamma_3}}{6(1-2\eta)}$  \\
 & (in the appendix)  & Stable: part ii in Figure \ref{fig:6} & & \\
 &  & Unstable: part iii in Figure \ref{fig:6} &  & \\
 &  & Stable spiral: part iv in Figure \ref{fig:6} & & \\
 & & Unstable spiral: part v in Figure \ref{fig:6} & & \\
 & & Centre: black curve in Figure \ref{fig:6} & & \\
\noalign{\smallskip}\hline
\end{tabular}
\end{center}
\end{table*}
For Table \ref{tab:7} we define the following functions
\begin{equation}
f_1(\eta) = \sqrt{\frac{3+12\eta-27\eta^2+3\sqrt{\eta}}{4(\eta-2\eta^2)}},
\end{equation}
\begin{equation}
f_2(\eta) = \sqrt{\frac{3+12\eta-27\eta^2-3\sqrt{\delta}}{4(\eta-2\eta^2)}},
\end{equation}
\begin{equation}
f_3(\eta) = \sqrt{\frac{3-12\eta+12\eta^2+6\sqrt{-\eta+5\eta^2-8\eta^3+4\eta^4}}{1-4\eta+4\eta^2}},
\end{equation}
where $\delta = 1-8\eta+30\eta^2-72\eta^3+81\eta^4$, furthermore $(\eta-2\eta^2)\neq 0$.\\

\begin{figure}[h]
\resizebox{.5\textwidth}{!}{%
  \includegraphics{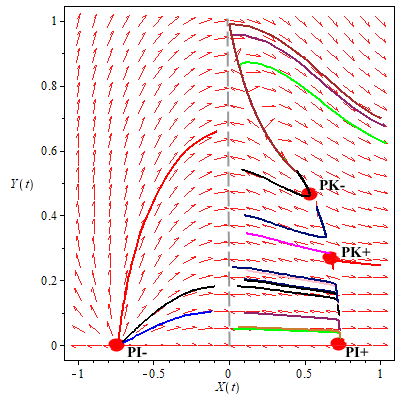}
  \includegraphics{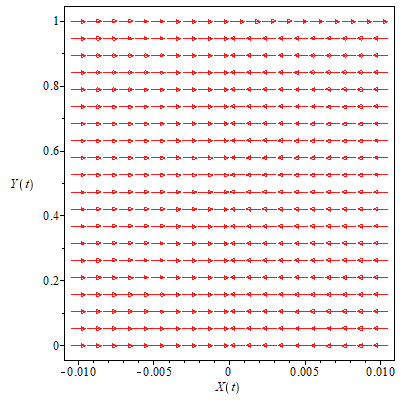} }
\resizebox{.5\textwidth}{!}{%
    \includegraphics{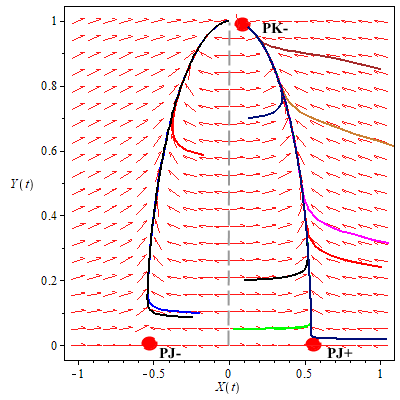}
      \includegraphics{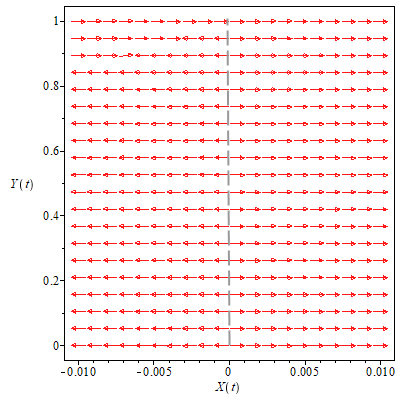} }
\caption{Phase portrait of the autonomous system of ordinary differential equations \eqref{eq:xprim} and \eqref{eq:yprim} for the case $\alpha=-\beta$ and the specific choice $(\beta = 2,\lambda=2.5)$ -top panels- and $(\beta =0.25,\lambda=1.75)$ - bottom panels-. The gray line corresponds to the separatrix $X=0$ which exists for $Y \neq 1$. It can be seen that the trajectories end at it for $\alpha/(1-\alpha+\beta) <0$ or depart from it if $\alpha/(1-\alpha+\beta)>0$. The behavior of the trajectories around the separatrix is shown in the right-hand side. In the upper scenario, Point $I_+$ and Point $I_-$ correspond to unstable nodes and Point $K_-$ denotes a saddle point. It is not relevant from cosmological considerations since there is no stable solutions. In the lower case, Point $J_+$ and Point $J_-$ behave as saddle and Point $K_-$ corresponds to an accelerated stable solution which can be the late time state of the universe. }
\label{fig:phase-space-alpha=-beta}
\end{figure}

The critical points of the dynamical system for the choice $\alpha= -\beta$, as well as their stability properties, are listed and briefly discussed below. For a couple of illustrative scenarios see Figure \ref{fig:phase-space-alpha=-beta}.\\

(i) Point $I_+$: Exists for $\eta \leq 0$, depending on the $\lambda$ value there is a saddle point or unstable point, so that their phenomenological properties remain the same independently of the potential. For the special case  $\eta = 0$ the universe is dominated by the scalar field kinetic energy $(X = 1, \Omega_m = 0$ and $Y=0)$. When $\eta$ is very large negative is also dominated by scalar field kinetic energy $(X \rightarrow 0.70$, $Y=0$ and $\Omega_m = 0.5$ . For this critical point there is no acceleration.\\

(ii) Point $I_-$:  Exists for non-positive $\eta$, depending on the $\lambda$ value there is a saddle point or unstable point, so that their phenomenological properties remain the same independently of the potential. The  scenario  $\eta=0$  the universe is dominated by the scalar field kinetic energy $(X = -1$, $Y=0$ and $\Omega_m = 0$). When $\eta$ is very large negative is also dominated by scalar field kinetic energy $(X \rightarrow -0.70$, $Y=0$ and $\Omega_m = 0.5$ . For this critical point there is no acceleration.\\\\

(iii) Point $J_+$: Exists for $\eta\geq 0$ ($\eta\neq \frac{1}{2}$), there is a saddle point, stable point or unstable point, depending on the $\lambda$ value. For the special case when $\eta = 0$ the universe is matter dominated $(X = 0$, $\Omega_m = 1$ and $Y = 0)$. For the case when $\eta$ tends to positive infinity we have $X\rightarrow 0.70$, $\Omega_m \rightarrow \frac{1}{2}$ and $Y=0$.
There is no acceleration for this critical point $J_+$.\\

(iv) Point $J_-$: Exists for $\eta \geq 0$  ($\eta\neq \frac{1}{2}$) there is a saddle point, stable point or unstable point, depending on the $\lambda$ value. The case $\eta = 0$ corresponds to a matter dominated solution $(X = 0$, $\Omega_m = 1$ and $Y = 0)$, and the limit $\eta \rightarrow \infty^+$ provides $X\rightarrow -0.70$, $Y=0$ and $\Omega_m \rightarrow \frac{1}{2}$. This solution is always decelerated.\\

(v) Point $K_+$: Exists for the region 5 reported in the appendix A. One part of this region is shown in Figure \ref{fig:5}. For $f_3(\eta)<\lambda<f_2(\eta)$ and $\eta\leq 0$ this is a saddle point. For $\lambda>f(\eta)$ and $\eta >1 $ this point is unstable and is very interesting to note that there is a region where is accelerated as can be seen in Figure \ref{fig:5}. \\

(vi) Point $K_-$: Exists for the region 6 reported in the appendix A and we only show a representative region in the Figure \ref{fig:6}. We see from Figure \ref{fig:6} that exists a saddle point, stable point, unstable point, spiral stable point, spiral unstable point and centre point for this critical solution depending on the values of $\lambda$ and $\eta$. The black dashed lines divide the accelerated and decelerated regions.\\

\section{Conclusion} \label{concl}
In this work we have performed a dynamical analysis in a spatially flat, homogeneous and isotropic spacetime for a nongravitational interaction scenario between pressureless dark matter and a quintessence scalar field with self-interacting exponential potential. Here we have considered a coupling function which is of cosmological interest, namely, $Q=- ( \alpha \dot{\rho_m} + \beta \dot{\rho_{\phi}})$, and we have studied the following scenarios $\alpha=0$, $\alpha = \beta$ and $\alpha = -\beta$, separately. We have found that this form of the interacting term can be rewritten as in equation \eqref{eq:finalinteraction}, so that its dependence on the matter density  and on the kinetic term of the scalar field is lineal and directly proportional to the Hubble parameter $Q = 3H \left[\frac{\alpha \rho_m + \beta \dot{\phi}^2}{1+\beta-\alpha}\right]$. In order to describe the cosmological evolution for each solution we have calculated various observables such as the effective equation of state parameter, the DE and DM dimensionless density parameters, the effective EoS parameter and the deceleration parameter. For every case, we have found the existence of singular points which can be related to relevant epochs in the history of the universe such as the matter dominated solution, the stiff matter universe, the scalar field dominated solution and the scaling scenario. Besides, it is necessary to mention a relevant aspect that has been overlooked in previous works \cite{Shahalam2015,Binayak,Bohmer2008}: the existence of a separatrix  whenever the coupling $\alpha$ is a non-vanishing constant. This completely modifies the structure of the phase portrait since divides it into two regions which are causally disconnected according to the sign of the kinetic term of the scalar field. In fact, at point $X=0$  whenever $Y_{0}\neq \pm 1$, the dynamical system fails to be continuously differentiable and therefore the system of cosmological equations does not satisfy the fundamental existence and uniqueness theorem for nonlinear ordinary differential equations. Therefore there can be two different trajectories with the same initial or final condition. This represents one of the main results of our analysis.  \\

In the case where the nongravitational interaction is absent $\alpha=\beta=0$, we recovered the results of standard quintessence. Moreover, when the energy transfer between dark components is present we found that some critical points may survive or they may completely disappear depending on the model parameters. We have found that except for the matter dominated solution, the remaining critical points correspond to scaling solutions where neither dark energy nor dark matter dominates. This represents a modification with respect to the non interacting scenario analyzed in \cite{wands} where the authors described the phase space of the universe with four critical points: (i) the ordinary dark matter dominated solution (always saddle), (ii) the kinetic energy dominated scenario described by a stiff fluid EoS which can be saddle or unstable, (iii)  the scalar field dominated solution representing a saddle or stable node and  (iv) the scaling solution which is always stable (node or spiral). \\

For the case $\alpha=0$ and $\beta \neq 0$, there are five critical points denoted by Point $O$, Point $D_+$, Point $D_-$, Point $E_+$ and Point $E_-$. Critical point $O$ exists for  $\beta \neq -1$ and behaves as unstable node or saddle. Point $D_+$ and Point $D_-$ have slightly more complicated stability conditions and denote as scaling solution whenever $0<\beta<1$. The limit $\beta \rightarrow 1$ corresponds to the matter dominated solution and the stiff  matter solution is recovered for $\beta=0$. This represents a modification to the mistaken results found in \cite{Shahalam2015,Binayak}, where the authors determine that these solutions always correspond to saddle points. Finally, although Point $E_+$ is always saddle, the true richness of the system is found at point $E_-$, since according to the values of the parameters $\beta$ and $\lambda$ the solution can be a stable node or a stable spiral, as well as an unstable node or unstable spiral o even a centre point for the case of one o more vanishing eigenvalues. It is important to mention that all solutions, except for Point $E_-$, are decelerating solutions.\\

The second scenario $\alpha = \beta$, possesses six critical points, namely Point $F_+$, Point $F_-$, Point $G_+$, Point $G_-$, Point $H^+$ and Point $H^-$. Here, Point $F_+$ and Point $F_-$ can be saddle or unstable and thus they cannot be the late-time state of the universe. These represent decelerating solutions where the quintessence tracks the dark matter behavior. The limit $Q \rightarrow 0$ corresponds to the ordinary stiff matter scenario. Point $G_+$ and  Point $G_-$ represent non-accelerating solutions which exist for non-negative $\eta$. For $\eta \neq 0$, the scaling solution behaves as stable node and then can be relevant at late-times. The matter dominated solution is retrieved in the limit $Q \rightarrow 0$ and it is always saddle. Point $H^+$ is always saddle and denotes only a transient epoch in the cosmological history. Finally, Point $H^-$ is the only accelerated solution and the stability properties indicate that this  can be a stable node or a stable spiral, as well as an unstable node or unstable spiral o even a centre point.\\

Finally, for $\alpha = - \beta$, there are six critical points, namely Point $I_+$, Point $I_-$, Point $J_+$, Point $J_-$, Point $K_+$ and Point $K_-$. Here, Point $I_+$ and Point $I_-$ can be saddle or unstable and thus they cannot be the late-time state of the universe. These represent decelerating solutions where the quintessence tracks the dark matter behavior. The limit $Q \rightarrow 0$ corresponds to the ordinary stiff matter scenario. Point $J_+$ and  Point $J_-$ represent non-accelerating solutions which exist for non-negative $\eta$. For $\eta \neq 0$, the scaling solution behaves as stable node and then can be relevant at late-times. The matter dominated solution is retrieved in the limit $Q \rightarrow 0$ and it is always saddle. Point $K_+$ is always saddle and denotes only a transient epoch in the cosmological history. Finally, Point $K_-$ is the only accelerated solution and the stability properties indicate that this  can be a stable node or a stable spiral, as well as an unstable node or unstable spiral o even a centre point.\\

We close this work by mentioning that, even though the existence of a nongravitational interaction between the dark energy and dark matter does not generate the appearance of new critical points, it does greatly modify the stability of the solutions at background level. There could still be the case that the interaction could leave their signatures on observables related to cosmological perturbations such as the density fluctuations and the power spectrum. Although such an investigation lies beyond the scope of the present paper, it could be interesting to investigate the relevance of this interacting scenario by confronting with cosmological observations.






\section{ACKNOWLEDGEMENTS}\label{ACKNOWLEDGMENTS}

The authors are grateful to FORDECYT-PRONACES-CONACYT for support of the present research under grant CF-MG-2558591 and CF-140630-UNAM-UMSNH. UN \\thanks the program Sistema Nacional de Investigadores (SNI) of the Consejo Nacional de Ciencia y Tecnolog\'{\i}a (CONACyT), the Programa para el Desarrollo Profesional Docente of the Secretar\'{\i}a de Educaci\'{o}n P\'{u}blica (PRODEP-SEP) of M\'{e}xico and the Coordinaci\'{o}n de la Investigaci\'{o}n Cient\'{\i}fica of the Universidad Michoacana de San Nicol\'as de Hidalgo (CIC-UMSNH) for financial support of his contribution to the present research. RDA also acknowledges CONACyT for the POSTDOCTORAL GRANTS CONACYT postdoc grant 350411 under which part of this work was performed. PP acknowledges CONACyT for grant 603730 and the Instituto de F\'isica y Matem\'aticas of the Universidad Michoacana de San Nicol\'as de Hidalgo for the support.


\appendix

\section{Regions of existence}
\label{apen1}
In this appendix we report the regions of existence of the critical points for different cases, where $\bigcup $ denotes "union"  and $ \bigcap $ corresponds to "intersection":\\
Region 1
\begin{eqnarray} \nonumber
&&\left( 0 \leq \beta< \frac{1}{3}\phantom{0} \bigcap \phantom{0} s_1(\beta)<\lambda \leq \sqrt{\frac{6}{1-\beta^2}}\right)\bigcup \\ \nonumber
&& \left( 0\leq \beta \leq \frac{1}{3}\phantom{0} \bigcap \phantom{0} \lambda = s_1(\beta)\right).
\end{eqnarray}
\\
Region 2
\begin{eqnarray}\nonumber
&&\left(-1<\beta\leq 0 \phantom{0}\bigcap \phantom{0} \lambda >0\right)\bigcup \\ \nonumber
&& \left(0<\beta\leq \frac{1}{3} \phantom{0}\bigcap \phantom{0} \lambda \geq s_1(\beta) \right) \bigcup \\ \nonumber
&&\left(\frac{1}{3}<\beta<1 \phantom{0}\bigcap \phantom{0} \lambda\geq \sqrt{\frac{6}{1-\beta^2}} \right).
\end{eqnarray}
where
\begin{equation}
s_1 = \frac{\sqrt{3+9\beta+6\beta^2+\sqrt{\beta(1+\beta)^3}}}{(1+\beta)}.
\end{equation}
\\
Region 3:
\begin{eqnarray} \nonumber
&&\left( \sqrt{3}<\lambda \leq \sqrt{6}\phantom{0}\bigcap \phantom{0} 0\leq \zeta <\frac{9-6\lambda^2+\lambda^4}{12\lambda^2}\right)\bigcup\\ \nonumber
&&\left( \sqrt{6}<\lambda<\sqrt{3+2\sqrt{3}}\phantom{0} \bigcap \right.\\ \nonumber
&&\left.\frac{6-\lambda^2}{3}+\frac{\sqrt{-54+45\lambda^2-12\lambda^4+\lambda^6}}{3\lambda}\leq \zeta<\frac{9-6\lambda^2+\lambda^4}{12\lambda^2} \right)\\ \nonumber
&&\bigcup \left(\sqrt{3}\leq \lambda \leq \sqrt{3+2\sqrt{3}}\phantom{0}\bigcap \phantom{0} \zeta = \frac{9-6\lambda^2+\lambda^4}{12\lambda^2} \right).
\end{eqnarray}
\\
Region 4:
\begin{eqnarray}\nonumber
&& \left( 0<\lambda <\sqrt{3}\phantom{0} \bigcap \phantom{0} -1<\zeta \leq 0\right)\bigcup \\ \nonumber
&&\left( \lambda = \sqrt{3} \phantom{0} \bigcap \phantom{0} -1 < \zeta < 0\right) \bigcup \\ \nonumber
&& \left(\sqrt{3}< \lambda \leq \sqrt{3+2\sqrt{3}} \phantom{0} \bigcap \phantom{0}-1<\zeta < \frac{9-6\lambda^2+\lambda^4}{12\lambda^2} \right) \\ \nonumber
&& \bigcup \left(\lambda> \sqrt{3+2\sqrt{3}}\phantom{0} \bigcap \right. \\ \nonumber
&& \left. -1<\zeta\leq \frac{6-\lambda^2}{3}+\frac{\sqrt{-54+45\lambda^2-12\lambda^4+\lambda^6}}{3\lambda}\right) \\ \nonumber
&& \bigcup \left(\sqrt{3}\leq \lambda \leq \sqrt{3+2\sqrt{3}} \phantom{0}\bigcap \phantom{0}\zeta = \frac{9-6\lambda^2+\lambda^4}{12\lambda^2} \right)
\end{eqnarray}
\\
Region 5:
\begin{eqnarray}\nonumber
&&\left(\eta \leq 0 \phantom{0}\bigcap \phantom{0}  f_3(\eta)\leq\lambda \leq f_2(\eta) \right)\\ \nonumber
&& \bigcup \left(\eta>1\phantom{0} \bigcap \phantom{0} \lambda \geq f_3(\eta)\right).
\end{eqnarray}
\\
where
\begin{equation}
f_3(\eta) = \sqrt{\frac{3-12\eta+12\eta^2+6\sqrt{-\eta+5\eta^2-8\eta^3+4\eta^4}}{1-4\eta+4\eta^2}},
\end{equation}
\begin{equation}
f_2(\eta) = \sqrt{\frac{3+12\eta-27\eta^2-3\sqrt{\delta}}{4(\eta-2\eta^2)}},
\end{equation}
and $\delta = 1-8\eta+30\eta^2-72\eta^3+81\eta^4$, furthermore $(\eta-2\eta^2)\neq 0$.
\\

Region 6:
\begin{eqnarray}\nonumber
&& \left(\eta<0 \phantom{0}\bigcap \phantom{0} \lambda>f_3(\eta)\right)\bigcup \left(\eta\leq 0\phantom{0}\bigcap \phantom{0} \lambda=f_3(\eta)\right) \\ \nonumber
&&\bigcup \left[\eta=0\phantom{0}\bigcap \phantom{0}\left(0<\lambda<\sqrt{3}\phantom{0}\bigcup \phantom{0} \lambda>\sqrt{3}\right) \right]\\ \nonumber
&&\bigcup \left( 0<\eta<\frac{1}{3}\phantom{0}\bigcap \phantom{0}0<\lambda\leq f_1(\eta)\right)\bigcup \\ \nonumber
&& \left(\frac{1}{3}\leq \eta<\frac{1}{2}\phantom{0}\bigcap \phantom{0}0< \lambda\leq f_2(\eta)\right)\bigcup\\ \nonumber
&& \left(\frac{1}{2}<\eta<1\phantom{0}\bigcap \phantom{0}0< \lambda\leq f_2(\eta) \right)\bigcup \\ \nonumber
&& \left(\eta =1 \phantom{0} \bigcap \phantom{0} \sqrt{3}<\lambda<\sqrt{3 \left(\sqrt{2}+1\right)}\right)\bigcup \\ \nonumber
&& \left(\eta >1 \phantom{0}\bigcap \phantom{0} f_3(\eta)< \lambda \leq f_2(\eta)\right)\bigcup\\
&& \left( \eta > 1 \phantom{0}\bigcap \phantom{0}\lambda = f_3(\eta) \right).
\end{eqnarray}

\section{Energy-momentum conservation}
\label{apen2}
In this Appendix we will  show the calculations around equations \eqref{eq:conservationttm} and \eqref{eq:conservationttphi}.
In the standard cosmological model dark matter and dark energy  are considered to be uncoupled  with separately conserved energy-momentum tensors. General covariance requires the conservation of their sum, so that
\begin{equation}
\nabla^{\nu} G_{\mu \nu} = \nabla^\nu (T_{\mu \nu}^m + T_{\mu \nu}^\phi) = 0.\label{ort}
\end{equation}The standard way of coupling two interacting matter components consists of adding a nonvanishing current $F_\mu = F_{\mu} (\rho_{m}, \rho_{\phi}, u^{\alpha}, \nabla_{\alpha} u^{\alpha}, \nabla_{\alpha} \rho_{m})$ to the right-hand side of the conservation equations, such that
\begin{equation}
\nabla^\nu T_{\mu \nu}^m = F_\mu, \,\, \nabla^\nu T_{\mu \nu}^\phi = - F_\mu. \label{derivada}
\end{equation}which guarantees the overall energy-momentum conservation. The projection onto the orthogonal 4-velocity $u^{\mu}$ defines the interacting term $Q$
\begin{equation}
u^{\mu} F_{\mu} \equiv  Q, \label{proy}
\end{equation}
where the four-velocity $u_\mu$ satisfies the condition $u^\mu u_\mu =-1$, and therefore $u^\nu \nabla_\mu u_\nu =0$. For a flat  Friedmann-Lema\^{i}tre-Robertson-Walker (FLRW) background spacetime  metric and a pressureless matter component, equations  \eqref{derivada} take the form
\begin{eqnarray} \label{estrella}
\dot{\rho}_{m} + 3 H \rho_{m} &=& Q, \\
(\ddot{\phi} + 3H \dot{\phi} + V_\phi ) \dot{\phi} &=& - Q.
\end{eqnarray}


\end{document}